\documentstyle[aps,prl,epsf,twocolumn,floats]{revtex}
\begin{document} 
\draft 
\twocolumn 
\title{Signature of Dynamical Localization in the Lifetime Distribution
of Wave-Chaotic Dielectric Resonators}
\author{Oleg A. Starykh$^{1}$, Philippe R. J. Jacquod$^{1}$, 
Evgenii E. Narimanov$^{2}$ and A. Douglas Stone$^{1}$} 
 
\address{$^{1}$ Department of Applied Physics, P.O.Box 208284, 
Yale University, New Haven, CT 06520-8284}
\address{$^{2}$ Bell Laboratories-- Lucent Technologies, 
700 Mountain Ave., Murray Hill NJ 07974} 

\date{ \today}

\maketitle

\begin{abstract}
We consider the effect of dynamical localization on the lifetimes
of the resonances in open wave-chaotic dielectric cavities. We
show that dynamical localization leads to a log-normal
distribution of the resonance lifetimes which scales with the
localization length in excellent agreement with the results of 
numerical calculations for open rough microcavities. 
\end{abstract}
\vspace*{-0.05 truein} 
\vspace*{-0.15 truein} 

The study of lifetime distributions
of finite quantum systems weakly coupled to a continuum 
is a subject of active experimental and theoretical investigation.
The nature of the spectrum of resonances depends strongly on the nature of 
the states of the finite system ``in isolation''. For example if those states 
are ergodically extended and 
structureless over the system then the resonances will show the behavior 
expected from 
random matrix theory, the famous Porter-Thomas distribution in the case of a 
single channel \cite{Porter-Thomas}.
A close relative of this resonance distribution has been measured in quantum 
dots in the  Coulomb blockade regime \cite{stone,Chang96}. 
More recently it has been pointed out 
that optical cavities with partially or fully chaotic ray dynamics would 
have interesting resonance properties and efforts have been made to 
characterize their distribution in various 
limits\cite{noeckel-stone,science,NHJS99}.  In a geometry which is 
approximately translationally invariant in one direction the wave 
equation becomes a scalar equation with 
a close formal analogy to the Schr\"odinger equation and the 
physics of the resonance spectrum 
becomes essentially the same for the optical and quantum systems. 
We will henceforth consider cylindrical dielectric resonators which are 
translationally invariant along their axis, but 
can be deformed in their cross-section.
The analogue of the classical limit of the Schr\"odinger equation is the 
limit of ray optics when the wavelength of the electromagnetic field 
is much shorter than the typical radius of the cavity $\lambda \ll R_0$. 
We will regularly use the term "quantum" to describe properties of the wave 
solutions which differ from the behavior of rays in the same geometry.
The motion of a light ray within the cavity is identical to that of a 
point mass in a classical 
billiard and the resulting bound states are the analogue of the eigenstates of 
"quantum billiards" \cite{stockman}. However, unless the index of 
refraction, $n$, is taken infinite, none of these states are truly bound, 
there always being some 
non-zero probability of escape from the cavity. Moreover in the case of 
a simple dielectric cavity the escape probability is strongly dependent on the 
angle of incidence of the ray.  
In particular, rays bouncing at the cavity's boundary with an angle 
of incidence $\chi$ smaller than the critical angle, $\chi_c= \sin^{-1}(1/n)$
(angles of incidence are defined from the normal to the boundary), 
are transmitted by refraction 
with high probability, while those with $\chi > \chi_c$ are trapped 
by total internal reflection, and can only escape with low probability 
by tunneling (evanescent leakage).
Semiclassically the (dimensionless) angular momentum of the ray in a circular 
cavity is 
$m=nkR_0 \sin \chi$, where $k=2\pi/\lambda$ is the wavevector (in vacuum) and 
$R_0$ is the radius of the cavity.  Hence a ray with angular 
momentum $m > kR_0$ will be 
strongly trapped whereas one with $m < kR_0$ will rapidly escape. 
Correspondingly, 
resonant states with mean values $\langle m \rangle > kR_0$ 
will have long lifetimes, whereas those with mean values less 
than $kR_0$ will have short 
lifetimes, i.e. there is a threshold value $m_c = kR_0$ for strong 
escape in {\it angular momentum space}.
In an undeformed (circular) cavity $m$ is an integral of motion and 
there are many 
exponentially long-lived ``whispering gallery'' resonances with $m > kR_0$.

For a generically deformed cavity angular momentum is not conserved, nor 
is there any other second constant of motion beyond the energy \cite{ellipse}. 
Hence the angular momentum can fluctuate.  The scale of those fluctuations 
depends on the existence of 
KAM tori in phase space which limit the diffusion in angle of incidence.   
Beyond some critical 
value of the deformation these barriers are destroyed and classical rays 
with initial
 angular momenta $m$ much larger than $m_c$ can now diffuse to arbitrarily 
low angular 
momentum and escape by refraction\cite{noeckel-stone}. As a result, 
even for $kR_0 \gg 1$
the lifetime of rays starting 
with $m > m_c$ is not exponentially long; the corresponding level width
$\Gamma_m$ can be estimated from the distance to the critical value in angular
momentum space: $\Gamma_m = D/(m-m_c)^2$.
(Here $D$ is the effective diffusion coefficient in
phase space, which in principle can depend on $m$.)  One might then guess 
that a cavity 
with such chaotic ray dynamics will no longer support any high-Q resonances.
However this is not necessarily the case, due to 
the phenomenon of "dynamical localization"
\cite{dyn_loc}.  It is now well-known that just as a 
random system exhibits exponential 
localization in real-space due to Anderson localization, 
the same kind of destructive 
interference can occur in a chaotic dynamical system, and 
suppress diffusion in the 
relevant phase space \cite{grempel}.

The condition for the onset of dynamical localization is that 
the diffusion time 
across the system be longer than the Heisenberg time defined by 
the inverse level 
spacing of the cavity: $t_H \sim \hbar \Delta^{-1}$.
For longer times than $t_H$ a wavepacket starts to ``resolve'' 
the discreteness of the 
spectrum and the spreading in angular momentum is suppressed. 
Based on an analogy with the kicked rotator \cite{fs}, 
the localization length $\xi$ is 
determined by the classical diffusion rate $D$, $\xi \sim D$.
Consider a state centered around an angular momentum $m_0$ such that
$m_0 - m_c \leq \xi$. In this case wavepackets can escape before 
their diffusion ceases 
and the classical picture is adequate.  Two different statistical 
behaviors are possible 
in this regime.  If the escape is determined by {\it slow } diffusion 
to a boundary 
where escape can occur the survival probability has been studied 
recently by several 
authors  \cite{borgonovi,maspero,masperoth} and they find characteristic 
power-law 
distributions.  Here the diffusion constant satisfies $L \ll D \ll L^2$, 
and it takes many collisions to cross the available phase space 
(for the optical cavities the role of the system size $L$ is 
played by $2nkR_0$, 
the number of angular momentum states available).
When the diffusion constant is larger, $D \sim L^2$, the motion is 
ballistic in the 
sense that the phase space is crossed in a few collisions; this 
situation leads to the 
Porter-Thomas distribution of resonance widths and the related 
distributions mentioned 
above \cite{Porter-Thomas,stone,alhassid,fyodor}.  However when 
dynamical localization 
dominates then
the lifetime of a localized state centered around angular momentum $m_0$
such that $m_0 - m_c \gg \xi$ becomes exponentially longer than the 
corresponding
classical diffusion time to the classical emission threshold.
Thus one has the possibility of high-Q resonances of completely non-classical, 
pseudo-random character, something not considered in the optics  literature to 
our knowledge (except in a very recent experiment in the microwave 
regime \cite{exper}).  
It therefore becomes of interest to understand the statistical 
distribution of resonance lifetimes in such a situation.

In the localized regime $\xi/L \ll 1$,
the angular momentum components of wavefunctions decay
exponentially away from their centers and one naturally expects exponentially
long average lifetimes for states centered far above the classical emission
threshold $ m_0 - m_c > \xi $. Recently N\"ockel and 
Stone \cite{noeckel-stone} compared the exact lifetimes of resonances of 
quadrupole-deformed microcavities with the mean classical diffusion time and 
found the lifetimes to be significantly longer in certain cases; they 
conjectured that these discrepancies arose from incipient dynamical 
localization. Indeed, dynamical localization has been shown to occur in 
certain closed cavities \cite{li,fs} and
a very recent experimental paper confirmed this phenomenon  
in microwave cavities of similar shape to those studied below \cite{exper}.  
However no detailed study has been made of the statistical and scaling 
properties of the lifetime in this regime.  These are the main topics of 
the current work. Below we will show that this localized regime is 
characterized by a very broad
(log-normal) lifetime distribution with scaling properties directly
related to the system's localization length $\xi$. 

\begin{figure}[t]
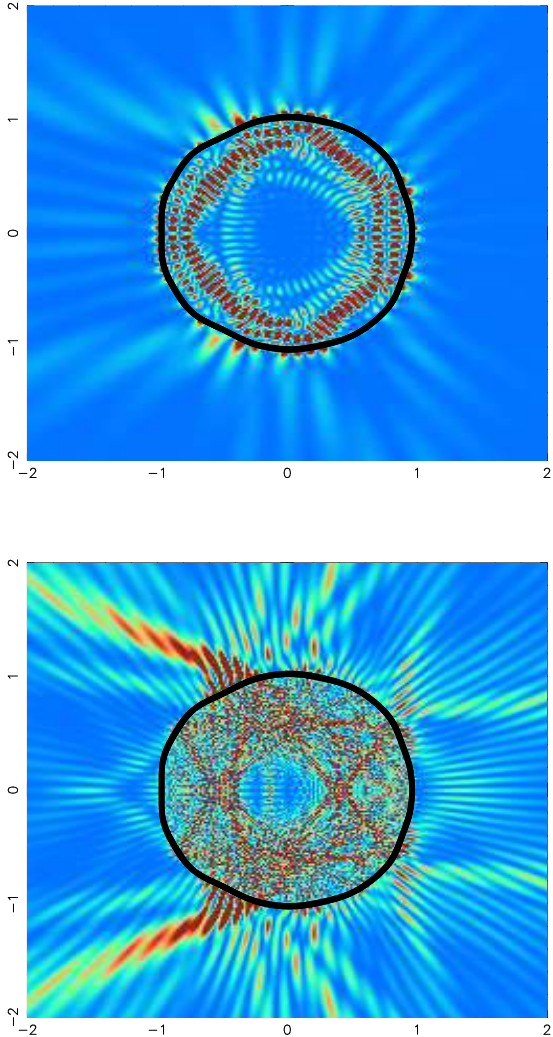

\includegraphics{state1.ps}
\includegraphics{state2.ps}
\vspace{14.5cm}
\caption{Top : Intensity plot of a resonance with $nkR_0=50$ in the rough
cavity with $\kappa=0.08$, $M=15$ and $n=2.5$. Red color corresponds
to the maximum of the intensity, and blue to the minimum.
Bottom : Intensity plot of a resonance with $nkR_0=150$ for the same set
of parameters as above.}
\label{fig:emission1}
\end{figure}

\begin{figure}[t]
\includegraphics{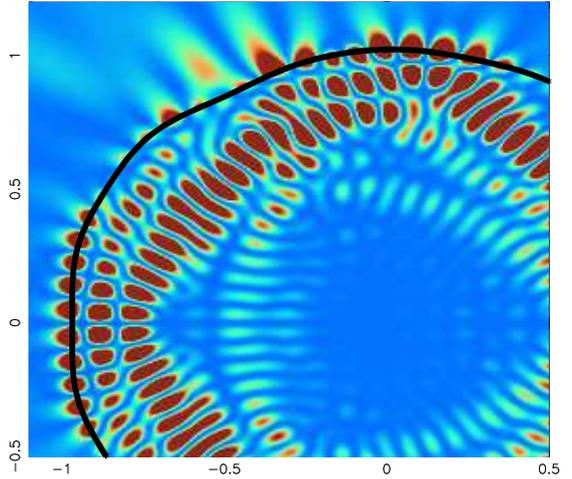}
\vspace{7.5cm}
\caption{Details of the wave intensity corresponding to the top resonance 
shown in Fig. \ref{fig:emission1}. The intricate structure of the wave
intensity is due to classically chaotic boundary scattering and makes the 
resonance clearly different from a standard whispering-gallery like high-$Q$
resonance.}
\label{fig:emission3}
\end{figure}

First, to illustrate the effect of dynamical localization on the
physical properties of the modes in the open cavity, we
show in Fig.\ref{fig:emission1} the 
real-space structure of two modes of a deformed cylindrical microcavity 
defined according to the model described immediately below. 
The two modes correspond to 
exactly the same shape of the cavity,
corresponding to fully ergodic classical ray dynamics, have the same average 
angular momentum $\langle m^2 \rangle^{1/2} \approx 0.5 nkR_0$, 
but differ in their wavevector and as a 
consequence (see below) differ in their localization lengths. As a result one 
resonance is in the quantum diffusive regime and the other in the localized 
regime. The qualitative difference is immediately apparent; the 
diffusive mode emits much 
more strongly and has a more dense spatial structure due to the large angular 
momentum spread in the state. The localized mode on the other hand emits 
weakly and appears to have a caustic similar to a regular whispering 
gallery mode, but a closer look at its spatial structure shows the pattern 
of nodes has an irregular character entirely different from the usual 
whispering gallery modes of circular 
resonators as can be seen on Fig. \ref{fig:emission3}.

We now define the model corresponding to Figure 1 and 2.
We consider an optically inactive, cylindrical microcavity with
an index of refraction $n>1$. The cross-section perpendicular to the 
cylinder's axis is given by a circle perturbed by $M$ harmonics of
random amplitude $-1/l \le a_\ell \le 1/l$, 
$R(\phi)=R_0[1 + \sum_{\ell=2}^M a_\ell \cos( \ell \phi)]$.
The average roughness of the surface is defined as 
$\bar{\kappa}=\sqrt{<\kappa^2(\phi)>_{\phi}}, ~\kappa(\phi)=(dR/d\phi)/R_0$.
This model was introduced by Frahm and Shepelyansky \cite{fs} with the 
condition of perfectly reflecting walls, and they referred to it as the
{\it rough} billiard  to contrast with the smoother quadrupolar deformations 
considered by N\"ockel and Stone \cite{noeckel-stone}. However the spatial 
wavelength of the roughness is still assumed to be large compared to the 
wavelength of the resonance. The advantage of a rough boundary is that 
the transition to classical chaos is achieved with much smaller amplitude 
of deformation making it easier to explore the parameter regime of fully 
chaotic classical motion and dynamically localized "quantum" behavior.  
As we shall see below, the open rough billiard has scaling and statistical 
properties essentially identical to a quasi-one-dimensional disordered system, 
whereas the quadrupole billiard does not.
For the rough billiard the 
classical dynamics can be well approximated by a discrete map
for which Chirikov's overlap criterion \cite{over} gives an
estimate of the critical roughness $\bar{\kappa}_c$ above which the classical
dynamics becomes fully chaotic as $\bar{\kappa}_c \sim M^{-5/2}$. 
The two deformation parameters $M$ and $\bar{\kappa}$ allow to reach a
classically fully ergodic regime characterized by a diffusion constant
(averaged over angular momentum) 
$D=\frac{4}{3}(\bar{\kappa}nkR_0)^2$ for $\bar{\kappa} \gg \bar{\kappa}_c$ and
quantum mechanically, one gets a localization length 
$\xi \sim D$ \cite{dyn_loc} so that the dynamically localized regime is
determined by $\bar{\kappa}^2 nkR_0 \ll 1$ \cite{fs}.
Keeping parameters of the cavity fixed and varying $kR_0$ one is able
to access states with greatly different localization lengths.

We restrict ourselves to the simplest case of TM-polarized electric field
$E({\bf r})$ parallel to the cylinder axis for which both the field
and its derivative are continuous at the cavity's boundary. 
This restriction is primarily for convenience, the TE modes obey a 
slightly different scalar equation which can be treated in a similar manner.  
It should be mentioned however that semiconductor quantum cascade 
micro-cylinder lasers studied in \cite{science}
emit solely in the TM mode due to a selection rule.
Maxwell's equation reduce then to a single scalar {\it {wave equation}},
$c^{-2}\partial_t^2 E({\bf r},t)=n^2({\bf r},\phi) \nabla_{\bf r}^2 
E({\bf r},t)$,
where the refraction index satisfies $n({\bf r})=n$ inside the cavity,
and $n({\bf r})=1$ outside.

We use the approach in which the resonances widths in wavevector are given 
by the imaginary part of the wavevector of the {\it quasibound} states 
defined by the following matching conditions.  First we expand the electric 
field in the angular momentum basis
($r \equiv \left| {\bf r} \right|$)
\begin{equation}
E\left({\bf r}, t\right) =
e^{- i c k t}
\sum_{m=-\infty}^\infty i^m  A_m\left(kr\right) e^{ i m \phi} ,
\label{psi_general}
\end{equation}
where
\begin{eqnarray}\label{coeff}
 A_m & = &
 \left\{
 \begin{array}{ll}
  \alpha_m  H_m^{+}\left(n k r\right)
  +  \beta_m H_m^{-}\left(n k r\right)  & {\rm if}\  r \leq R(\phi) ,\\
  \gamma_m H_m^{+}\left(k r\right)  & {\rm otherwise}.
  \end{array}
  \right.
  \end{eqnarray}

This expansion corresponds to the so-called {\it Siegert} boundary 
conditions \cite{siegert} in which the states
have only an {\it outgoing} component  at infinity ($H^{\pm}_m(x)$
are Hankel functions of first and second type, respectively). Such boundary 
conditions cannot be satisfied for real $k$ and are only satisfied 
for discrete complex $k$.  It can be shown that the imaginary part 
of the $k$ values which satisfy Eq. (2) are the poles of the 
true unitary (on-shell) S-matrix of 
the scattering problem. From the expansion coefficients in Eq. (\ref{coeff}) 
we define vectors $\vec{\alpha}$,
$\vec{\beta}$ and $\vec{\gamma}$. The fields inside and outside the cavity 
are related by the continuity of the field and its derivative on the boundary 
and (after integration around the boundary) one of these equations can be 
used to eliminate $\vec{\gamma}$ leaving a linear relation between 
$\vec{\alpha}$ and $\vec{\beta}$.  The matrix expressing this relation we 
call $\tilde{\cal{S}}$.  Moreover the regularity of the field at the origin 
$r=0$ implies $\vec{\alpha}=\vec{\beta}$, thus a secular equation for the 
resonant values of $k$ is obtained of the form:

\begin{equation}
\tilde{\cal{S}}\vec{\alpha}=\vec{\alpha}.
\label{eq:quant} 
\end{equation}
We use the notation $\tilde{\cal{S}}$-matrix because in the case of 
a closed billiard 
the matrix so-defined is actually the unitary S-matrix of the scattering 
problem of a wave incident outside the impenetrable 
billiard \cite{doron,dietz}.  
In our case the matrix so-defined is non-unitary for any $n<\infty$
and for real $k$, the eigenvalues of $\tilde{\cal{S}}$ have the form
$\lambda_r=\exp(i(\varphi_r + i\delta_r))$, where 
both $\varphi_r$ and $\delta_r>0$ are real functions of momentum $k$.
The subscript $r$ numbers states in a deformed cavity where angular
momentum is not conserved.
Exact quantization of the cavity - solving Eq. (\ref{eq:quant}) exactly - 
implies $\lambda_r = 1$ so that
the exact implementation of this procedure requires finding
a complex $k = q - i\gamma $ such that 
$\varphi_r(q - i\gamma)=\delta_r(q-i\gamma)=0$.
The corresponding inverse lifetime $\Gamma_r$ is then given by
$c$ times $\gamma$ via the dispersion relation,
$\Gamma_r=c\gamma$.
However approximate lifetimes can be found by a much  more efficient 
procedure which is 
extremely helpful if one wishes to study full distributions as we do here.
First, it is straightforward to show \cite{unpub} that 
when $\gamma $ is small, 
simply finding the complex eigenphases 
$\lambda_r=\exp(i(\varphi_r + i\delta_r))$ for real
$k$ determines the imaginary part of $k$ on resonance by the relation:
$\gamma = - \delta/\varphi^{\prime}$ where $\varphi^{\prime}$ is the 
derivative of the 
real part of the phase with respect to momentum for real $k$.  
Moreover, since this derivative
can be shown to be slowly-varying on the scale of the 
level spacing $\Delta k$, it is
not necessary even to quantize the real part of the phase 
(i.e. to find the real $k$ which makes $\varphi (k) = 2 \pi 
\times {\rm integer}$).  
The function $\varphi^{\prime}$ can be easily 
calculated for the circular cylinder and this relation and the assumption of 
slow variation of the derivative can be confirmed explicitly 
(for the case where $\gamma$ is small).
Therefore we can generate large lifetime ensembles simply by diagonalizing 
the matrix $\tilde{\cal{S}}$ for real $k$ and extracting the imaginary 
phase $\delta$, by which means we
generate $\sim 2nkR_0$ inverse lifetimes per diagonalization.
This procedure is motivated by the work of Doron and Frischat who noted 
that the statistical properties of closed billiards changed little away 
from the exact quantization condition (in their case it was the distribution 
of splittings of semiclassically degenerate states) \cite{doron}.  
The linear relation between $\delta_r$ and $\gamma_r$ has been
independently proposed earlier by Hackenbroich \cite{gregor} and 
demonstrated for the case of the circle.

To best relate the localization properties of the eigenstates, 
which apply to a closed cavity, to the distribution of lifetimes in 
an open cavity, we employ a perturbative formalism which was recently 
developed specifically to treat open optical resonators \cite{NHJS99} 
(it is similar in spirit to well-known quantum perturbative scattering 
approaches such as R-matrix theory in the single-level approximation).
According to that theory
narrow resonance widths $\Gamma \ll \Delta$ ($\Delta$ is the resonance
spacing) can be computed from the expectation value of an antihermitian
operator $V$ taken over eigenstates $\left. \right| \alpha^{(0)}
\rangle$ of the matrix ${\cal M}$, which describes some effective
\lq\lq closed cavity\rq\rq

\begin{eqnarray}
\Gamma & = & b_0 \langle \alpha^{(0)} \left| V \right| \alpha^{(0)}
\rangle  = b_0
\sum_{m,m'} \alpha_m^{(0)*} V_{mm'} \alpha_{m'}^{(0)}
\label{eq:gamma_pert} 
\end{eqnarray}
Explicitely, 
\begin{equation}
{\cal M}=(J'J') -(1/n) ( J' {H^{+}}') 
\left( H^{-} H^{+} \right)^{-1} \left( H^{-} J \right), 
\label{M}
\end{equation}
$V$ is the antihermitian part of ${\cal M}$, and
the matrix elements $(\bar{Z} Z)$ are defined as
\begin{eqnarray} 
\left( \bar{Z} Z \right)_{ \ell m } = 
\frac{i^{m-\ell}}{2 \pi} 
\int d\phi \ 
\bar{Z}_{\ell} \left(k  \right) 
Z_m\left(k\right) \ e^{ i \left(m - \ell\right) \phi } , 
\label{eq:me} 
\end{eqnarray} 
$Z_m(k)$ and $\bar{Z}_m(k)$ stand for either 
$H^{\pm}_m\left(k R\left(\phi\right)\right)$, the Bessel function 
$J_m \left(n k R\left(\phi\right)\right)$, or their derivative. 
The coefficient $b_0$ in (\ref{eq:gamma_pert}) depends only on the 
hermitian part ${\cal H}_0$ of ${\cal M}$, i.e. it is determined by the
properties of the \lq\lq closed system\rq\rq , 
$b_0^{-1}= \langle \alpha^{(0)} \left|\partial {\cal H}_0/\partial k \right| 
\alpha^{(0)}\rangle
\approx \sum_m  \alpha_m^{(0)*} (J' J'')_{mm'} \alpha_{m'}^{(0)}$,
and can be regarded as a 
normalization factor. 
An eigenstate, localized at angular momentum 
$m_0 \gg m_c \approx k R_0$, will have exponentially small width $\Gamma$.
Therefore, for such a resonance in the semiclassical limit 
$\Gamma \ll \Delta$, and Eq. (\ref{eq:gamma_pert}) is appropriate.
More details on this formalism can be found
in \onlinecite{NHJS99}.

For an exponentially localized state 
one generally has
\begin{eqnarray}
\left| \alpha_m \right| 
\sim \exp\left( - \frac{\left| m - m_0\right|}{ 2 \xi} \right)
\label{eq:alpha}
\end{eqnarray}

\begin{figure}[t]
\epsfxsize=7.5cm
\epsfbox{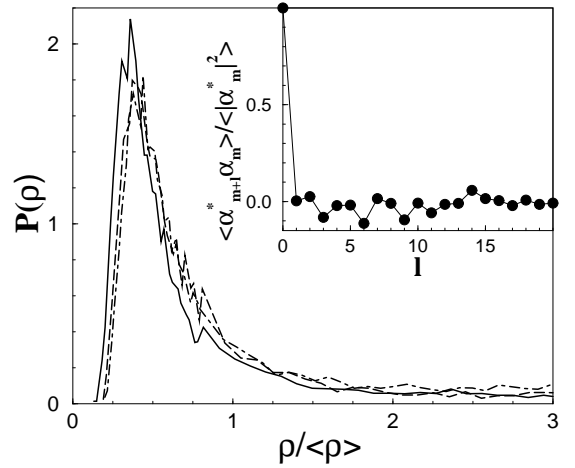}
\vspace{3mm}
\caption{Distribution of the inverse participation ratio $\rho$
for $n=3$, $M=15$, and 
$(nkR_0,\bar{\kappa})=(150,0.02)$ (solid line), $(100,0.03)$ (dashed line) 
and $(50,0.06)$ (long-dashed line). $\bar{\kappa} nkR$ has been kept constant,
which result in a stable average IPR $\langle \rho \rangle = 0.1 \pm 0.01$.
Inset: 
Normalized correlation function (\ref{eq:corr}) for $nkR_0 =100$,
and $\bar{\kappa}=0.03$.}
\label{fig:ipr}
\end{figure}

We consider the regime of large localization length 
$\xi \sim {\bar{\kappa}^2} k^2 R_0^2 \gg 1$.
Since the rough billiard is classically chaotic,
the {\it phases} of the coefficients $\alpha_m$ change rapidly with the
angular momentum, and it is natural to assume 
that its correlation function satisfies

\begin{eqnarray}
\langle \alpha^*_{\ell + m} \alpha_\ell \rangle_\ell = \delta_{m,0} 
\langle \left| \alpha_\ell \right|^2 \rangle_\ell
\label{eq:corr}
\end{eqnarray}
where the average is performed over an angular momentum interval
$l \in \left[m_0-\delta \ell/2,m_0+\delta \ell/2 \right]$ such that 
$1 \ll \delta \ell < \xi$. This behavior is
illustrated in the inset to Fig. \ref{fig:ipr} for one typical set 
of cavity parameters, corroborating the validity of 
the assumption (\ref{eq:corr}).

As follows from Eq. (\ref{eq:me}) and the definition of $V$,
the matrix elements
$V_{m m'}$ in angular momentum representation vary on a
scale of $\delta m \sim \bar{\kappa} k R_0 \gg 1$, and therefore
we can replace the product $\alpha_m^* \alpha_{m'}$ in
Eq. (\ref{eq:gamma_pert}) by its
average value $\langle \alpha_m^* \alpha_{m'} \rangle$ over the 
interval $\left| m - m'\right| \sim \bar{\kappa} k R_0$.
Together with (\ref{eq:corr}) this leads to the diagonal
approximation
\begin{eqnarray}
\langle \alpha_0 \left| V \right| \alpha_0 \rangle 
\approx \sum_{|m| = kR_0}^{nkR_0} \left| \alpha_m \right|^2 V_{mm}
\label{eq:diag}
\end{eqnarray} 
The matrix element 
\begin{equation}
V_{mm}=-\frac{1}{2 n}[(J' {H^{+}}')_{m\ell}
(H^{-}H^{+})^{-1}_{\ell\ell'} (H^{-}J)_{\ell' m} - c.c.]
\end{equation}
includes both the refractive (classical) escape
from the resonator (for $|\ell|,|\ell'| < m_c = kR_0$) and the \lq\lq tunneling
escape \rq\rq (corresponding to evanescent leakage \cite{noeckel-stone}, 
for $|\ell|,|\ell'| > m_c$). 
To evaluate the sum over angular momenta we use the stationary phase-based
technique, developed in \cite{splittings} in the context of the calculation
of level splittings in rough billiard.
The \lq\lq classical\rq\rq refraction contribution is found to be
\begin{equation}
\Gamma_{m_0 > m_c}^{\text{class}} \approx \frac{k^2 R_0}{n\xi}
\exp\left( - \frac{ m_0 - kR_0}{\xi}\right).
\label{eq:gamma_sim}
\end{equation}
This result shows that  an exponentially-long lifetime can be due to the
exponentially-small wavefunction component leaking outside the
classically totally-internally-reflected region.  To see if this process controls the lifetime
we need to compare this result with the direct \lq\lq tunneling escape\rq\rq
 contribution to the lifetime. The latter process involves angular
momenta only above emission threshold $m_c$. 
For an estimate, it is then sufficient to evaluate the
linewidth of the state above $m_c$ in the circular cavity, which can
be thought of as a state with zero localization length. We find
that the tunneling contribution is also exponentially small ($n \gg 1$),
\begin{eqnarray}
\Gamma_{m_0 > m_c}^{\text{tunn}} =&& -(1/n){\rm Im} \left[H^{+}_{m_0-1}(kR_0)/
H^{+}_{m_0}(kR_0)\right] \nonumber\\
\approx &&  \exp\Big(2\sqrt{m_0^2 - m_c^2} +
2m_0\ln(m_c) \nonumber\\
&& - 2m_0\ln(m_0 + \sqrt{m_0^2 -m_c^2})\Big).
\label{eq:tunn}
\end{eqnarray}
Competition between the classical escape [Eq.(\ref{eq:gamma_sim})] and
the tunneling [Eq.(\ref{eq:tunn})] is strongest for $(m_0 - m_c)/m_c \ll 1$, 
i.e. when
the width of the tunneling barrier is smallest.
Comparing the two contributions in this region we find that 
$\Gamma^{\text{class}} \gg
\Gamma^{\text{tunn}}$ for $\xi \gg \sqrt{kR_0}$, and 
$\Gamma^{\text{tunn}} \gg \Gamma^{\text{class}}$ in the opposite limit.
Restriction on the range of $\xi$ weakens as one moves to higher
angular momentum, and for $(m_0 - m_c)/m_c \sim 1$ the classical escape
mechanism 
always dominates over the tunneling one.
The tunneling escape therefore
 is relevant only for very small deformations $\kappa \ll 
\left( k R_0 \right)^{-3/4}$, which produce short localization length.
Thus, lifetime of the states with very short (essentially zero)
localization length is determined by the tunneling escape, whereas that
of the more extended states with $\xi \gg 1$ by the classical 
emission from their exponentially weak tails at $m_c = kR_0$.

Having established a relation between the lifetime and the localization 
length of a corresponding closed cavity, the distribution of lifetimes 
then follows from the 
distribution of localization lengths.
In one-dimensional and quasi-one-dimensional disordered systems it is now 
well-established that the distribution of the inverse localization length is 
typically normally distributed around an
average $1/\xi_0$ \cite{gauss}.  Moreover Frahm and Shepelyansky 
have explicitly shown \cite{fs} that the problem of the rough 
billiard maps onto a variant of the kicked rotor problem and hence 
to an ensemble of band random matrices \cite{ipr}, which also describe 
quasi-1d disordered systems.  Here the angular momentum index plays the 
role of the site coordinates in the disordered systems, with an ideal 
lead (the continuum) accessible at $m_c=kR_0$. Collisions with the rough 
boundary correspond to random hopping between sites, which are at most 
$\bar{\kappa} nkR_0$ lattice spacings apart \cite{fs}.
Once the critical angular momentum, corresponding to
the classical emission border $m_c$, is reached, the wavepacket escapes the
system  and never returns.  Hence we may assume that the inverse localization 
length in our problem has an approximately normal distribution around its mean 
(our ensemble here is of boundary realizations)
\begin{eqnarray}
P(\xi) \sim \exp[-\frac{(1/\xi - 1/\xi_0)^2}{2\sigma^2}],
\label{eq:pxi}
\end{eqnarray}

Therefore, as follows from (\ref{eq:gamma_sim}) and (\ref{eq:pxi}),
the resonance lifetime is distributed log-normally,

\begin{equation}
P(t) d(\ln(t)) \sim \exp[-\frac{\ln^2(t/t_0)}{2\sigma^2 (m_0 - kR_0)^2}] 
d(\ln(t))
\label{eq:log-normal}
\end{equation}
where

\begin{equation}
\ln(t_0) = (m_0 - kR_0)/\xi_0
\label{eq:logtime}
\end{equation}
and the derivation is done in a leading
logarithm approximation, so that the pre-exponential factor  
in (\ref{eq:gamma_sim}) is neglected.  This result is then very natural: 
the  distribution of lifetimes in open dynamically-localized cavities is 
log-normal for the resonances localized far from the classical emission 
threshold  $m_0 \gg kR_0$.  This is entirely analogous to
the conductance distribution of localized chains, which will be log-normal 
for a fixed distance from the ends \cite{gauss}
(see \cite{texier} for log-normal distribution of delay times/resonance
widths). We note that the relationship between 
dynamical localization and Anderson localization was first placed on firm 
footing in a seminal paper by Fishman, Grempel and Prange \cite{grempel}.

\begin{figure}[t]
\epsfxsize=7.5cm
\epsfbox{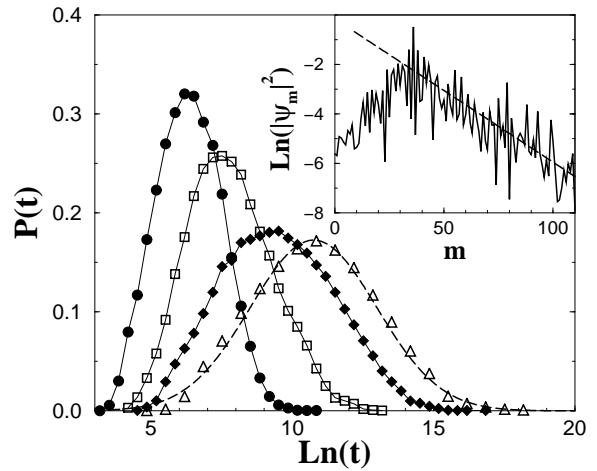}
\vspace{3mm}
\caption{Lifetime distribution for $n=3.5$, $nkR \approx 110$,
$\bar{\kappa} = 0.018$ and $M=15$ for states with average angular momentum
$m_0/(nkR_0)=0.5$ (circles), $0.6$ (squares), $0.7$ (diamonds) and
$0.8$ (triangles). This last distribution is fitted according to
Eq. (\ref{eq:log-normal}) with $\sigma \approx 0.04$ (dashed line).
The emission border is at $m_c \approx 0.29 nkR$. Each distribution
is constructed from 8000 to 13000 lifetimes obtained from 
1013 boundary realizations of the rough cavity.
Inset : Typical localized eigenstate for the same parameter set.
The dashed line indicates an exponential decay corresponding to a
 localization length of $\xi = 16$.}
\label{fig:distribution}
\end{figure}

Equation  (\ref{eq:log-normal}) essentially
relies on the two assumptions: first , Eq. (\ref{eq:corr}) that the phase
of the wavefunction components are randomly distributed with
no long-range correlations and second, that the eigenstates are exponentially
localized with a normal distribution of localization lengths.
We now test the validity of these two assumptions for our rough
microcavities. The validity of Eq. (\ref{eq:corr})
is confirmed by the sharp drop of the correlation function
$\langle \alpha^*_{\ell + m} \alpha_\ell \rangle_\ell/ 
\langle \left| \alpha_\ell \right|^2 \rangle_\ell$ for $m>0$ which is 
clearly seen from the numerical results presented 
in the inset to Fig. \ref{fig:ipr}.
The localization properties are investigated
by computing the distribution of the Inverse Participation Ratio (IPR)
defined as $\rho=\sum_m |\alpha_m|^4/\sum_m |\alpha_m|^2$.
The IPR measures the inverse number of effective eigenstates components and
thus allows to distinguish between localized and delocalized 
states. Generally, in the localized phase $\langle \rho \rangle$ is independent
of the system size since at most $\xi \ll L$
sites contribute to the sum and $\langle \rho \rangle \sim \xi^{-1}$. 
In the other limit of ergodic states, all sites
contribute equally and $\langle \rho \rangle \sim L^{-1}$ in this case.
Between these two limits, a variety of behaviors may occur depending
on the inner structure of the eigenstates.
In Fig. \ref{fig:ipr} we show IPR distributions for three different
parameters sets corresponding to the same average localization length
$\xi \sim (\bar{\kappa}nkR_0)^2 \equiv 9$. The three distributions
are indeed stable under parameter variations keeping $\xi$
constant and this shows that not only
the average IPR/localization length \cite{fs}, but also the full IPR 
distribution obeys a one-parameter scaling with $\bar{\kappa}nkR$. 
The situations is very similar to the one studied in Ref. \cite{ipr} 
for BRM with a band width $1 \ll b \approx \sqrt{2 \xi}$ for which
the IPR distribution was analytically computed. Similar deviations
from \cite{ipr} as seen on Fig. \ref{fig:ipr} are also present for 
BRM with not too large band widths, so that these numerical results confirm
the universality of the dynamically localized regime, quite analogous
to quasi-one-dimensional disordered systems. We also 
illustrate this exponential localization by showing one typical state in the
inset to Fig. \ref{fig:distribution}. 

Having tested the validity of the main assumptions on which
(\ref{eq:log-normal}) relies, we present in Fig. \ref{fig:distribution}
distribution of lifetimes for the classically-chaotic, dynamically
localized regime of the open rough microcavity. The distributions shown
correspond to resonances
centered in intervals of width $\delta m/(nkR_0) = 0.1$ around angular momenta
$m_0/(nkR_0)=0.5$, 0.6, 0.7 and 0.8, well above the classical
threshold $m_c/(nkR_0)\approx0.29$. Clearly, the distributions are log-normal
and their widths increase as one moves away from the classical
emission border $m_c=kR_0$.
Furthermore, the agreement with Eq. (\ref{eq:log-normal})
is quantitatively confirmed by a direct fit of the broadest of 
these distributions (see dashed line on Fig.\ref{fig:distribution}). 

\begin{figure}[t]
\epsfxsize=7.5cm
\epsfbox{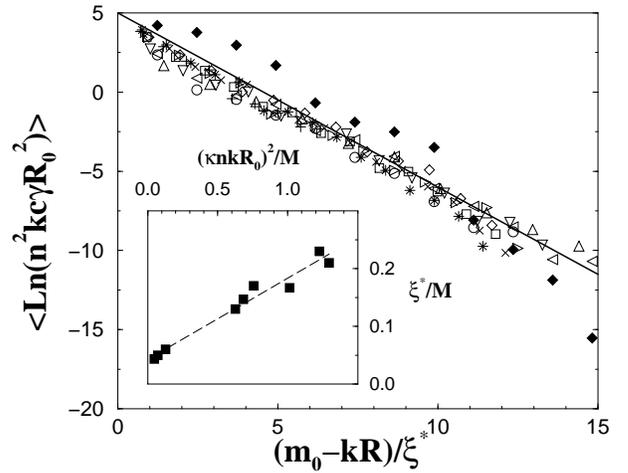}
\vspace{3mm}
\caption{Scaling of the average logarithm of lifetime
$\langle\ln t_0\rangle$ vs. $(m_0 - kR)/\xi^*$ for parameters
$95 <nkR <195$, $2.5<n<5$, $0.008< \bar{\kappa} <0.03$ and $10<M<30$.
Open symbols correspond to the rough deformation for which a one-parameter
scaling exists. The full diamonds correspond to a quadrupolar billiard
($M=2$) for which the diffusion is affected by classical
invariant structures in phase space.
Inset : Localization length as extracted from the scaling shown on the
main pannel.
}
\label{fig:scaling}
\end{figure}

We expect by analogy to the scaling theory of localization that the 
logarithmic average of  the lifetime will exhibit a universal scaling 
behavior.  This expectation is confirmed by the data shown in  
Fig. \ref{fig:scaling} where we present numerical results for the
scaling obeyed by ${\rm ln} t_0$. Log-averaged lifetimes
for different parameter sets have been computed for at least 2000 lifetimes
in narrow energy windows $\delta m/(nkR_0) = 0.2$ around given angular momentum
$m_0 > m_c$ for different values of indices of refraction $1.5 \le n \le 4$, 
wavelengths $75 \le nkR_0 \le 180$ and roughnesses $ \bar{\kappa}$. All
presented results are in the localized regimes $\xi < m_0-m_c$
and the corresponding curves have been put on top of each other by
a one-parameter scaling.
Fig. \ref{fig:scaling} demonstrates the validity of the linear 
relation (\ref{eq:logtime}) as is indicated by the straight line. 
The exact parameter dependence of the scaling can be deduced from the 
analogy to the scaling theory of localization.  In this case the 
Thouless conductance $g=\Gamma/\Delta \epsilon$ is the scaling quantity 
(or its logarithm in the localized regime); here $\Gamma$ is the 
resonance width, 
and $\Delta \epsilon$ is the mean level spacing.  In our case 
$\Gamma = c \gamma$ (where $\gamma$ is the width in momentum space), 
but $\Delta \epsilon $ differs from the corresponding Schr\"odinger equation, 
due to the different
dispersion relation for the wave equation.  
Taking this into account one 
finds that the analogue of the dimensionless conductance 
is $g \sim  n^2k c \gamma R_0^2$.
and it is the logarithm of this 
dimensionless quantity which we plot against $(m_0 - kR)/\xi^*$.
Fig. \ref{fig:scaling} allows to identify 
the scaling parameter $\xi^*$ with the localization length up to a free
parameter.  That this scaling holds for the rough cavity demonstrates that 
localization length
is independent on the angular momentum as is expected for an homogeneously
diffusive system. The situation is fundamentally different for 
a quadrupolar cavity ($M=2$) as can be seen on Fig.\ref{fig:scaling} (see black
diamonds). Obviously one scaling parameter is not sufficient to bring
the corresponding curve on top of the other 
ones satisfying (\ref{eq:logtime}). This indicates an angular 
momentum dependent
diffusion constant which directly follows from the effective local map
derived for this particular case \cite{jensthesis}.
Furthermore, in the regime corresponding to the presented data 
for the quadrupolar deformation, small invariant torii and islands of stability
still survive, resulting in strongly localized wavefunctions
with short localization length determined essentially by the size
of remaining classical structures. Because of this, and unlike the 
situation in the rough cavity, lifetime of such states is determined
by the tunneling escape (see discussion after Eq.(\ref{eq:tunn})).
Therefore a clean demonstration of exponential dynamical localization is 
difficult in the quadrupolar billiard.

Further confirmation that the extracted scaling parameter is indeed related 
to the system's localization properties is given in the inset to 
Fig. \ref{fig:scaling} where $\xi^*$ is plotted against 
the diffusion constant $D = (\bar{\kappa} nkR_0)^2$ 
as derived from the effective rough map \cite{fs}.
This inset gives an unambiguous confirmation of the above derived 
relation between localization length and log-averaged lifetime. Note that
$\xi^*$ has a linear dependence on the diffusion constant $D$
even at small $D$ where the relation $\xi \sim D$ does not hold.
For large $D$ however, the relation $\xi^* \sim \xi$ holds.
 
To summarize, we have presented the first study 
of the lifetime distributions of a quantum-chaotic 
open system in the dynamically localized regime. This study was 
greatly expedited by the linear relation between the
complex phases of the eigenvalues of the nonunitary scattering
matrix away from exact quantization and the imaginary part of the 
corresponding exactly quantized complex wavevector, which
 has allowed us to generate sufficiently large lifetime statistics to
demonstrate the log-normal form of their distribution. The lifetime 
distribution has been derived analytically assuming a normal distribution
of inverse localization lengths and phase randomness of the wavefunctions, 
using a recently developed semiclassical method, the usefulness of which 
is thus further demonstrated \cite{NHJS99}, and confirmed numerically.
The log-normal distribution is a hallmark of localized disordered systems and 
hence our results deepen the analogy between dynamical and Anderson 
localization and point out an optical observable which can in principle 
be measured to demonstrate this distribution.
The possibility of high-Q resonances in deformed rough cavities 
(which are nonetheless smooth on the scale of the wavelength) should 
be of interest in optical studies of scattering from small particles, 
however their random nature seem to make such resonances unsuitable 
for applications.

We have benefitted from interesting discussions with F. Borgonovi,
C. Texier and Y. Fyodorov and would like to thank G. Maspero for sending us 
his
thesis and A. D. Mirlin for communicating us several interesting references. 
We acknowledge the support of the Swiss NSF and the NSF grant No. PHY9612200.


\begin{references}
\bibitem{why_need_many} 
An exponentially large number of lifetimes is needed to
construct a log-normal distribution function.
\end{references}
\end{document}